\documentstyle[11pt,revpasp,epsf]{article}
\def\edcomment#1{\iffalse\marginpar{\raggedright\sl#1\/}\else\relax\fi}
\reversemarginpar

\def\kms{\relax \ifmmode {\,\rm km\,s}^{-1}\else \,km\,s$^{-1}$\fi}
\def\ha{\relax \ifmmode {\rm H}\alpha\else H$\alpha$\fi}
\def\hb{\relax \ifmmode {\rm H}\beta\else H$\beta$\fi}
\def\hi{\relax \ifmmode {\rm H\,{\sc i}}\else H\,{\sc i}\fi}
\def\hii{\relax \ifmmode {\rm H\,{\sc ii}}\else H\,{\sc ii}\fi}
\def\h2{\relax \ifmmode {\rm H}_2\else H$_2$\fi}
\def\lha{\relax \ifmmode L_{{\rm H}\alpha}\else $L_{{\rm H}\alpha}$\fi}
\def\shi{\relax \ifmmode \sigma_{{\rm HI}}\else $\sigma_{\rm HI}$\fi}
\def\sh2{\relax \ifmmode \sigma_{{\rm H}_2}\else $\sigma_{{\rm H}_2}$\fi}
\def\degr{\hbox{$^\circ$}}
\def\arcmin{\hbox{$^\prime$}}
\def\arcsec{\hbox{$^{\prime\prime}$}}
\def\deg{\hbox{$^\circ$}}

\def\sec{\hbox{$^{\prime\prime}$}}
\def\fdg{\hbox{$.\!\!^\circ$}}
\def\fs{\hbox{$.\!\!^{\rm s}$}}
\def\farcm{\hbox{$.\mkern-4mu^\prime$}}
\def\farcs{\hbox{$.\!\!^{\prime\prime}$}}
\def\degd#1.#2{ #1\fdg#2 }                 % degrees over decimal point
\def\mind#1.#2{ #1\farcm#2 }               % minutes over decimal point
\def\secd#1.#2{ #1\farcs#2 }               % seconds over decimal point
\def\hhh{\ifmmode {\rm ^h}              % hours symbol
         \else {${\rm ^h}$}
         \fi}
\def\sss{\ifmmode {\rm ^s}              % seconds symbol
         \else {${\rm ^s}$}
         \fi}
\def\hms#1h#2m#3s{                      % hms format (for RA)
                  \relax
                  \ifmmode #1^{\rm h}\,#2^{\rm m}\,#3^{\rm s}
                  \else \hbox{$#1^{\rm h}\,#2^{\rm m}\,#3^{\rm s}$}
                  \fi
                 }
\def\dms#1d#2m#3s{                      % dms format (for Dec)
                  \relax
                  #1\degr\,#2\arcmin\,#3\arcsec
                 }
\def\hmsd#1h#2m#3.#4s{                  % hms format with decimal point (RA)
                      \relax
                      \ifmmode #1^{\rm h}\,#2^{\rm m}\,#3\fs#4
                      \else \hbox{$#1^{\rm h}\,#2^{\rm m}\,#3\fs#4$}
                      \fi
                     }
\def\dmsd#1d#2m#3.#4s{                  % dms format with decimal point (Dec)
                      \relax
                      #1\degr\,#2\arcmin\,#3\farcs#4
                     }

\begin{document}
\title{Observations of Barred Galaxies}
\author{Johan H. Knapen}
\affil{University of Hertfordshire, Department of Physical Sciences,
Hatfield, AL10 9AB, U.K.} 

\begin{abstract}

We review general observational properties of bars in galaxies, and
relations between bars, their dynamics, and (circum)nuclear activity.
We consider new measurements of bar fractions and of the distribution of
bars with host type, bar strength, and bar pattern speeds.  Bars
redistribute material radially, leading to flattened abundance gradients
and a diminished degree of disk-wide two-fold symmetry in the
distribution of star-forming regions.  We discuss recent results on
statistical correlations between bars and AGNs in samples of active and
non-active galaxies, and on circumnuclear regions of star formation in
the cores of barred galaxies.  Finally, we review the limited, but
promising, work published to date on bar fractions at cosmological
distances, and list a number of issues where further observational
progress should be expected.

\end{abstract}

\section{Introduction}

Bars are common structures in disk galaxies. Just over half of all
galaxies are barred (Sellwood \& Wilkinson 1993; Sect.~2), a fraction
that only goes up slightly when near-infrared (NIR) rather than optical
images are used for the classification, contrary to claims that the use
of NIR imaging substantially increases bar fractions. The main
observational characteristics of bars in the optical and NIR are an
elongated light distribution, pairs of curved dust lanes which show up
in broad-band or color index maps, and characteristic isophote twists.
Radio and millimeter observations often show gas concentrations in the
central regions, with some emission along the dust lanes.  Kinematic
evidence for gas streaming along the main body of the bar can be seen in
\hi\ or millimeter line emission (e.g. Bosma 1981; Knapen et al. 1993).

Gas response to the barred potential is delayed, creating offset leading
shocks inside the corotation radius. Enhancements in the gas surface
density are observed where such shocks occur. Generally, this is
believed to be where the dust lanes are observed in the optical/NIR,
although Beck et al. (1999) claim from polarimetric radio continuum
observations that the shock fronts are significantly offset from the
dust lanes in the bar of NGC~1097, the first galaxy for which such data
could be obtained.  An important consequence of the elongation of the
dominant stellar orbits and (given the high stellar mass fraction) of
the mass distribution, is that the gravitational potential of a barred
galaxy is non-axisymmetric. Through shocks in the gas which dissipate
energy, such a potential naturally provides a mechanism for angular
momentum loss in inflowing gaseous material. Bars can thus be expected
to channel gas from the disk of a galaxy to the central kpc regions
(e.g. Shlosman, Begelman \& Frank 1990; Athanassoula 1992; Sellwood \&
Moore 1999).  Channeling of gas by a hierarchy of stellar and gaseous
bars (Shlosman, Frank \& Begelman 1989) is frequently invoked to explain
the occurrence of central or circumnuclear starburst, or AGN activity
(see Sect.~4).

The astrophysical importance of bars lies both in their common
occurrence in disk galaxies, and in their ability to facilitate gas
inflow. In this paper, recent observational results on selected aspects
of barred galaxies are reviewed. Related theoretically or
observationally oriented reviews include those by Sellwood \& Wilkinson
(1993) which concentrates on the dynamics of barred galaxies, by Buta \&
Combes (1996) on ring galaxies, and several reviews in the proceedings
of the 1995 Alabama meeting (e.g. Elmegreen 1996; Freeman 1996; Roy
1996). In the present volume, Friedli (1999) reviews the birth, aging
and death of bars, whereas Shlosman (1999) discusses the dynamical
evolution of central regions of disk galaxies. This review is more
observationally biased and complements those by Friedli and Shlosman.

This paper discusses selected basic observational properties of bars
where recent progress has been reported in the literature (Sect.~2),
radial mixing due to bars (Sect.~3), and the relation of bars to
Seyfert or circumnuclear star formation activity (Sect.~4 and 5). Bar
fractions at cosmological distances are discussed in Sect.~6.  The
summary (Sect.~7) includes a brief list of open questions and ideas
for future work.

\section{Basic Bar Properties}

\subsection{Are All Galaxies Barred?}

How common are bars in galaxies? And does this bar fraction go up,
possibly to 100\%, if observations are considered at NIR wavelengths
and/or very high resolution? In other words, are all galaxies barred at
some level? Although new bars are being found in well-known galaxies
such as NGC~1068 (NIR imaging by Thronson et al. 1989) or Centaurus A,
for which Mirabel et al. (1999) claim that the bisymmetric structure
observed in dust emission at mid-IR and submillimeter wavelengths is
indicative of a gaseous bar, it is far more difficult to obtain reliable
statistics that can help answer the questions raised above.

Sellwood \& Wilkinson (1993) compiled statistics on bar fractions using
data from three major catalogues, and found that whereas the bar (SB)
fraction was roughly constant across catalogues and morphological types,
at 25\%-35\% of disk galaxies, the intermediate (SAB) fraction was as
high as 25\% in the RC2 (de Vaucouleurs et al. 1976), but substantially
lower in the other two catalogues analysed.

\begin{figure}
\vspace{1cm}
\plottwo{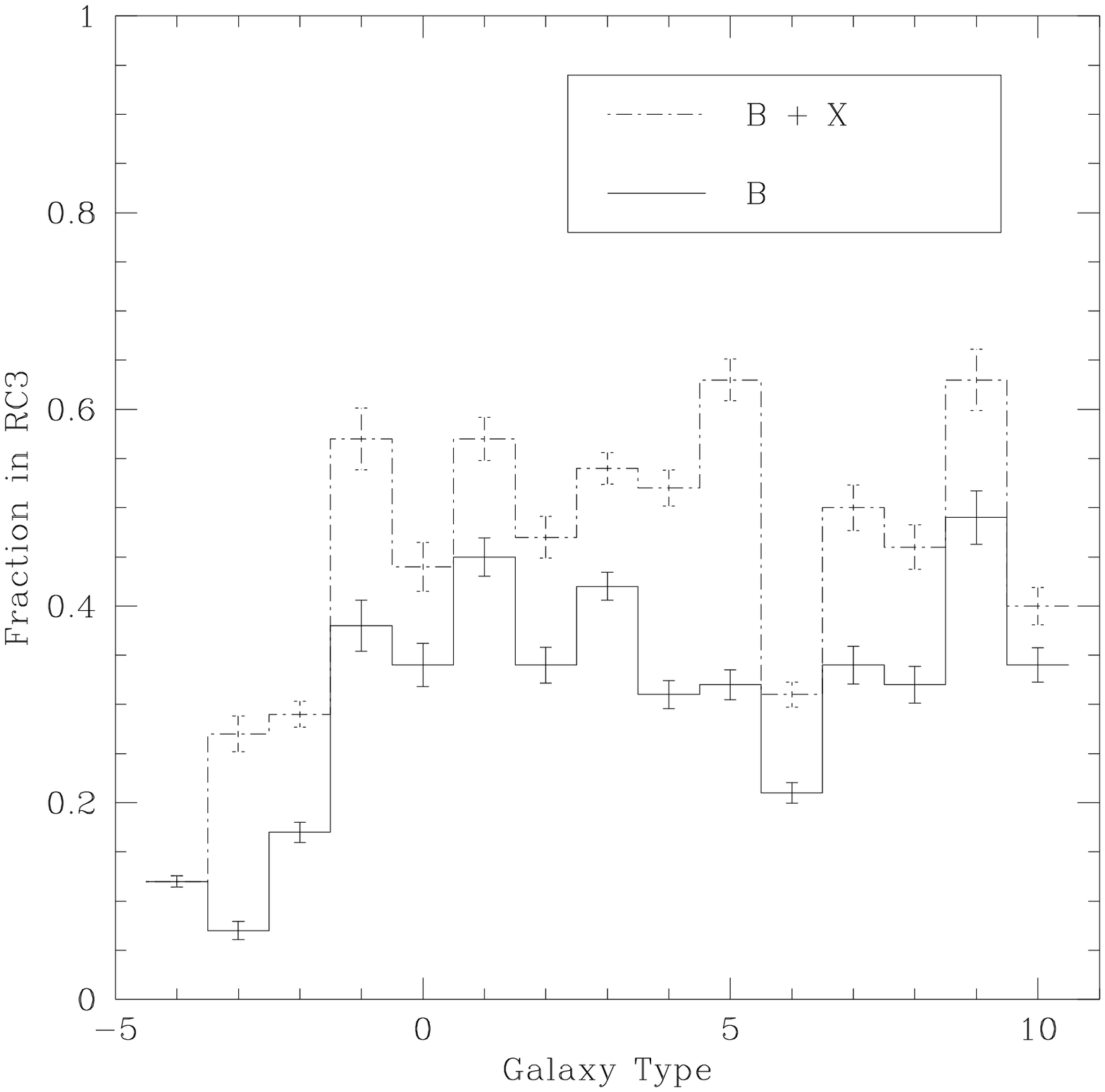}{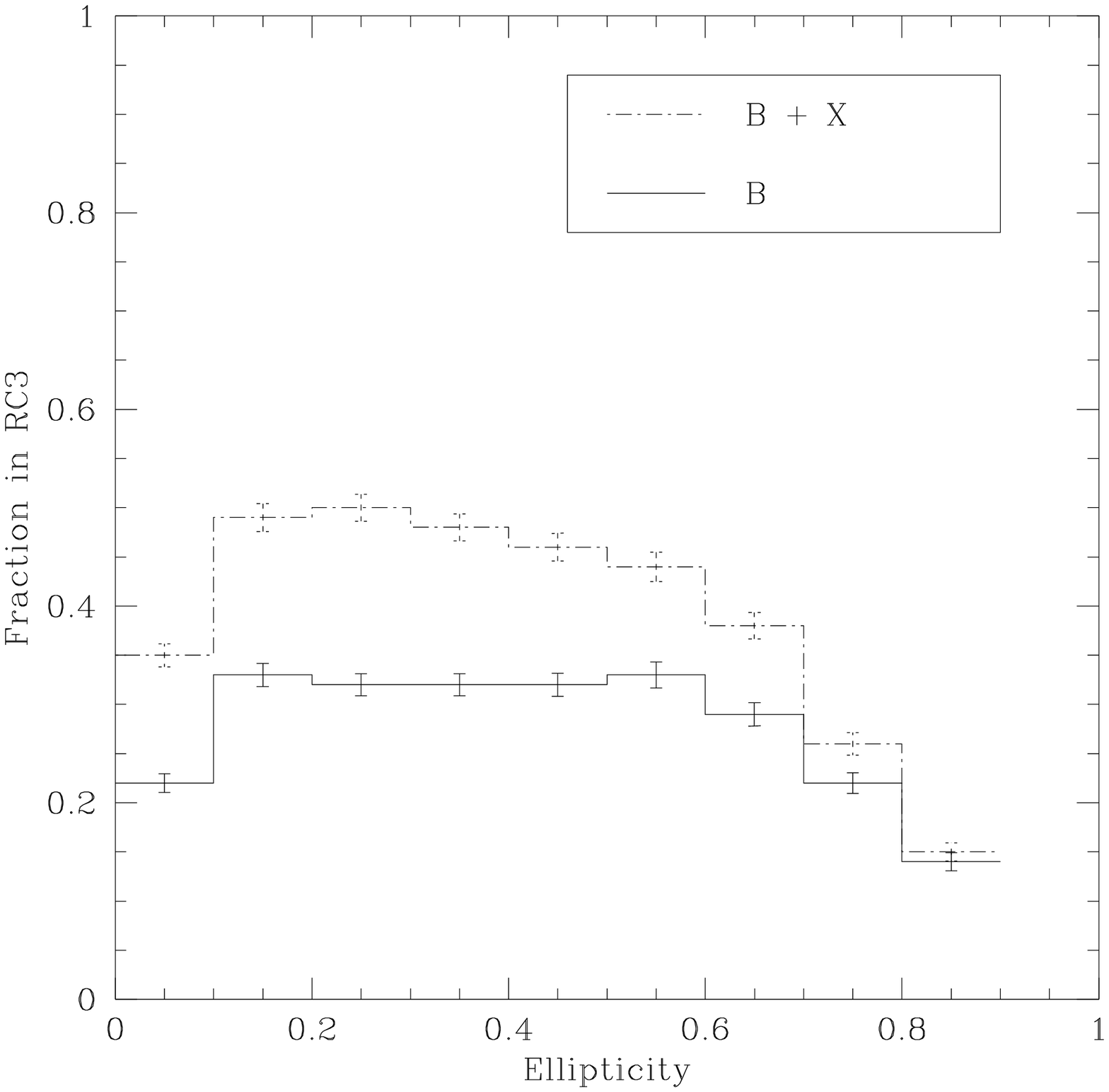}
\caption{Bar fractions as determined from the morphological
classifications in the RC3 as a function of morphological type (left
panel) and ellipticity $(1-b/a$; right panel). B and B$+$X fractions 
are shown separately. Data from Knapen et al (1999a).} 
\end{figure}

Studying the distribution of morphological classification in the RC3
catalogue (de Vaucouleurs et al. 1991), Knapen, Shlosman \& Peletier
(1999a) confirm an overall bar fraction (including intermediate type, X)
of between 50\% and 60\%. Fig.~1 shows this bar fraction as a function
of morphological type and ellipticity (or inclination) of the host
galaxies. Several interesting trends can be noted. (1) About 35\% of all
galaxies are classified as B, or barred, and this fraction is remarkably
constant across morphological types ranging from type=-1 (S0) to type=10
(Sm). (2) The total barred fraction (B$+$X) is roughly constant at a
level between 50\% and 60\%, again over the full range of types. This
also shows that the ratio B/X, strong to intermediate bars, is constant
(at about 2). (3) The B fraction is constant with ellipticity up to
$\epsilon\sim0.6$, whereas the X fraction is constant up to
$\epsilon\sim0.4$. This effect can be interpreted simply as one of
selection: it is progressively more difficult to distinguish bars from
morphology alone in more inclined galaxies, and this effect is more
noticeable for less well-defined, presumably weaker, bars.

So between one half and two thirds of all disk galaxies contain a
stronger or weaker bar, as based on the classification in the RC3, with
about twice as many galaxies being classified as B than as X.

This determination is  based on morphological classification
from optical plates exposed in blue or visual light, which deliver
galaxy images with a resolution of a few arcsec, and with often
overexposed central regions. Being dominated by an older stellar
population, in contrast to e.g. spiral arms, bars are most readily
detected in the NIR. It is thus somewhat surprising that the bar
fraction in spirals goes up by only about $10\%-15\%$ when NIR digital
images are used for the classification, even when these are at subarcsec
resolution (e.g. Mulchaey \& Regan 1998; Knapen et al. 1999a). These
studies are as yet based on small samples of objects as compared to the
RC3, and results from more extensive NIR surveys or even complete sky
surveys should be used for confirmation.

Certainly not all spirals are barred in the conventional
sense. Small-scale bars, nested within primary bars or not, are
discovered at high resolution in the NIR or using adaptive optics
techniques (e.g. Regan \& Mulchaey 1999; Erwin \& Sparke 1999; Combes et
al., in preparation), but not in all galaxies (e.g. Laine et al. 1999a;
Regan \& Mulchaey 1999). However, small-scale oval distortions,
miniature bars on scales smaller than tens of parsec, or
non-axisymmetries due to structures other than bars or ovals,
e.g. spiral arms, may have some or all of the dynamical effects ascribed
to bars, but would not normally end up in statistics on bar fractions.

\subsection{Bar Strength}

Whereas the parameter bar strength can be defined rather easily
theoretically, it is hard to determine observationally, and is usually
measured as, e.g., axis ratio or length of the bar.  Sandage \& Tammann
(1987) and de Vaucouleurs et al.  (1991) catalogue barred galaxies as
either ``AB'' or ``X'' for intermediate bars, or ``B'' for truly barred
galaxies, but these classifications are not strictly delimited by bar
strength as measured more quantitatively from digital images.  Several
quantities have been used as indications of bar strength, e.g.  the
fraction of total luminosity in the bar (e.g.  Sellwood \& Wilkinson
1993), or the length of the bar relative to the total galaxy size.

Martin (1995) used the simple but very useful measure of the so-called
apparent bar ellipticity.  In equivalence with the usual classification
of elliptical galaxies, this parameter is defined as $\epsilon_{\rm
b}=10(1-b/a)$, where $b/a$ is the apparent ratio between bar minor and
major axes.  Thus a galaxy without a bar will have $\epsilon_{\rm b}=0$,
one with a bar with a high axis ratio, presumed to be a strong bar,
$\epsilon_{\rm b}=8$.  Martin (1995) studied the lengths and axis ratios
of the bars in a sizable sample of galaxies, and found, among other
results, that bars in early-type galaxies are longer on average.  Martin
also confirmed the previously reported (e.g.  Athanassoula \& Martinet
1980) relation between bar length and bulge diameter, but a large
scatter completely hides any correlation of bar axis ratio, or strength,
with morphological type. 

Some interesting correlations with bar `strength' have appeared in the
literature over the past few years. For example, high star formation
activity seems to occur in galaxies with strong bars (but not all
strongly barred galaxies have a high star formation rate; Martin 1995;
Martinet \& Friedli 1997). Aguerri, Beckman \& Prieto (1998) find a very
weak relation between bar strength and the location of the co-rotation
radius ($R_{\rm CR}/R_{\rm bar}$), which remains to be confirmed.

\subsection{Bar Pattern Speed}

The pattern speed of a bar is one of its most defining and important
properties, and its determination is critical for a detailed study of
bar dynamics. Unfortunately, measuring bar pattern speeds remains a
difficult task in practice. Elmegreen (1996) reviewed observations of
bar pattern speeds and their implications, and this section will only
discuss more recently published results.

In general, there are three main ways to measure the pattern speed of a
bar. Firstly, with data from high signal-to-noise long-slit spectra of
bars, the Tremaine-Weinberg method (based on the continuity equation)
can be used for stellar bars (Tremaine \& Weinberg 1984). Gerssen,
Kuijken \& Merrifield (1999) used the method to obtain a value for the
pattern speed for the early-type barred galaxy NGC~4596 (SBa), placing
the co-rotation radius just outside the bar ($R_{\rm CR}=1.2R_{\rm
bar}$).

Secondly, morphological information can be used to deduce the location
of resonances, which in combination with a rotation curve leads to a
pattern speed determination. Fourier techniques are often used in this
respect. Aguerri et al. (1998) performed a Fourier analysis of optical
images of 10 barred galaxies and found $\bar{R}_{\rm CR}=1.2\bar{R}_{\rm
bar}$. Using morphology in another sense, Elmegreen, Wilcots \& Pisano
(1998) studied streaming motions from an \hi\ map of the SABd galaxy NGC
925 and found the remarkably large value of $R_{\rm CR}=3R_{\rm
bar}$. Canzian (1998) studied the extent of spiral structure and relates
it to the location of the main resonances for the unprecedented number
of 109 galaxies, including almost 60 barred spirals, and about 40 ringed
spirals. A result of interest to the present discussion is that Canzian
plots the distribution of a parameter describing the extent of the
spiral, and finds a peak for barred and for ringed spirals at a position
indicating that the corotation resonance occurs at or just outside the
end of the bar. This is under the assumption that bar and spiral are
resonantly related, implying that the spirals can only extend between
corotation and the outer Lindblad resonance. Canzian finds a
considerable number of barred spirals where the spiral extends further
into the disk than expected in this picture, which may indicate that
weak bars do not quite end near corotation, or that bars and spirals are
not resonantly related in all galaxies.

Thirdly, one can construct families of numerical models for a galaxy
with varying pattern speeds, and from comparison of the observed and
modeled morphology certain pattern speed values can be excluded and a
preferred value determined.  Lindblad, Lindblad \& Athanassoula (1996),
for example, compare their numerical gas simulations with morphological
observations of NGC 1365 (SBb), and find $R_{\rm CR}=1.2-1.3R_{\rm
bar}$. For NGC~1300 (also SBb), Lindblad \& Kristen (1996) find that two
models represent the galaxy morphology, with $R_{\rm CR}=1.3$ and
$2.4R_{\rm bar}$. Laine, Shlosman \& Heller (1998) find $R_{\rm
CR}=1.1R_{\rm bar}$ for NGC~7479 (SBbc), in agreement with the value
obtained previously from modeling and morphology by Sempere, Combes \&
Casoli (1995).

The main problem with pattern speeds remains that it is very hard
observationally to determine them. Several methods are now regularly
used in the literature, but the error analysis is often not convincing,
and individual results must be critically examined.  In spite of this,
the consensus seems to converge, with rather few exceptions, toward
$R_{\rm CR}=1.1-1.2R_{\rm bar}$, or bars ending just before the
co-rotation radius. This result continues to confirm the bar-aligned
orbit theory (e.g.  Contopoulos 1980). There are no significant trends
of pattern speed with e.g. bar strength or morphological type of the
host galaxy.

\section{Radial Mixing Induced by Bars}

\subsection{Abundance Gradients}

Several studies published over the past decade have provided ample
evidence that radial abundance gradients are flatter in barred than in
non-barred galaxies, as reviewed by Roy (1996). Martin \& Roy (1994)
even found a trend of stronger bars having flatter gradients.  Flatter
abundance gradients in barred galaxies have been interpreted as evidence
for radial mixing in the disk, induced by the bar.

In a recent study, Dutil \& Roy (1999) extend the study of abundance
gradients to early-type galaxies. This has only now been done because
abundance gradients are usually measured from line ratios in individual
\hii\ regions, which are most plentiful and most easily observed in
later type galaxies. Dutil \& Roy find that the abundance gradients in
early type galaxies are shallower than those in non-barred late-type
galaxies, but similar to strongly barred late-type galaxies. They
interpret this in terms of late-type non-barred galaxies (Sd, Sc, Sbc)
developing a strong bar, and evolving via SBb and SBc types into
earlier-type (Sa, Sb) galaxies. Theoretical arguments can be found to
support such an evolution (e.g. Martinet 1995).

\subsection{Symmetry in Star Forming Regions}

Rozas, Knapen \& Beckman (1998; see also Knapen 1992) studied the
two-fold symmetry of star-forming regions along the two main spiral arms
in a dozen galaxies, and find an anti-correlation between symmetry and
bar strength. The sample galaxies are all grand-design spirals, and are
thus two-fold symmetric in their spiral arm shapes, but it is the
symmetry of the ``pearls on the string'', the \hii\ regions along the
spiral arms, that is considered here. An example of strong symmetry is
M51, where all main star-forming complexes can be identified with a
counterpart some 180\deg\ away in azimuth but at the same galactocentric
radius, and whose locations can be identified with the locations of
density wave resonances (Knapen et al. 1992). Measuring this degree of
two-fold symmetry quantitatively and relating it to the presence and
strength of the large-scale bar of the host galaxy, led Rozas et
al. (1998) to their conclusion that apparently bars destroy global
symmetry in the distribution of star-forming regions.

The anti-correlation between bar strength and degree of symmetry of the
star formation regions within two-armed spirals is interpreted along the
same lines as the abundance gradient results, namely as a result of
radial mixing in barred galaxies, which destroys the disk-wide symmetry
in the location of star forming regions which was present originally as
a result of the global density wave.

The sample used by Rozas et al. (1998) is small, but if their
conclusion on bars and symmetry can be confirmed generally, the 
technique might be used for detecting bars through H$\alpha$ or FUV
imaging, possibly to substantial redshifts because the bar itself,
mosty devoid of recent star formation, would not have to be detected.

\section{Bars and Nuclear Activity}

Since theoretical arguments indicate that bars are efficient vehicles
for gaseous inflow from the disk (Shlosman et al. 1990; Athanassoula
1992), gas accumulation can be expected in the central regions of barred
galaxies. The inflow is slowed down towards the center (more so in the
presence of inner Lindblad resonances -- ILRs), and nuclear star-forming
rings and disks can be formed. Massive nuclear disks and nuclear rings
are themselves subject to non-axisymmetric dynamical instabilities which
drive the gas further in by means of gravitational torques, leading to
the formation of nested bars and fueling stellar and nonstellar activity
in the center (Shlosman et al. 1989). As a result, a direct
observational relation between the occurrence of a bar, and of
(circum)nuclear activity might be expected. Gaseous nuclear bars are
expected to be short-lived, probably around $10^7$ yrs, and are
therefore challenging to detect.  Besides, their observational
characteristics are as yet completely unknown.

Previous observational work suggests strongly that central starburst
hosts are preferentially barred (e.g.  Heckman 1980; Balzano 1983;
Devereux 1987), but the issue has been much more controversial for
Seyfert galaxies. After early claims of higher bar fractions in
Seyferts (e.g. Adams 1977; Simkin, Su \& Schwarz 1980) had been
criticized (e.g., Balick and Heckman 1982) for not having a sufficiently
well-defined control sample, only recently has more work appeared on the
subject.

Ho, Filippenko \& Sargent (1997) compare the morphology as obtained from
the RC3 of active galaxies, including Seyferts and LINERs, as identified
from their spectra, with that of non-active galaxies. Although the
source of their classifications (the RC3, thus optical plate material)
can be criticised, their active and control samples were well
matched (unlike, e.g., those used by McLeod \& Rieke 1995; Moles,
M\'arquez \& P\'erez 1995; or Hunt \& Malkan 1999). Ho et
al. conclude that the bar fractions in AGN and non-active galaxies are
equal. Using new NIR images of Seyferts and of a matched control sample,
Mulchaey \& Regan (1997) determine bar fractions, and although they find
that more galaxies are barred than in the Ho et al. study (as expected
through the use of NIR array imaging), they find no evidence for
enhanced bar fractions in their active sample.

Peletier et al. (1999) obtained NIR $K$-band images at subarcsecond
resolution of the complete CfA sample of Seyferts, and of a well-matched
control sample. From a morphological analysis of these data, Knapen et
al. (1999a) find that whereas about 60\% of their non-active galaxies
are barred, almost 80\% of the Seyfert galaxies are barred (after
excluding from the analysis those galaxies that are too small to study
their morphology, interacting, or edge-on). Given that the statistical
uncertainties in these numbers are about 8\%, the significance of this
result, that slightly more Seyfert hosts are barred, though by no means
all, is at the $2.5\sigma$ level, and needs further confirmation using
larger samples. The differences between the Seyfert bar fractions as
found by Mulchaey \& Regan (1997) and Knapen et al. (1999) can be
explained by the higher spatial resolution of the latter study
($\sim\secd 0.7$ vs. $\sim\secd 1.5$), and the strict use of a set of
well-defined criteria to determine the presence of a bar.

It is possible that a direct correlation between bars and nuclear
activity might exist for bars at smaller scales, but this remains to be
confirmed (the study by Regan \& Mulchaey 1999, who used HST NIR and
optical imaging to see whether small bars occur particularly often in
Seyfert cores, can only be regarded as a first step due to the limited
sample size and to the use of ad hoc criteria for the presence of
small-scale bars, assuming that they have the same morphological
characteristics as large-scale bars). Miniature spiral arms seem to be
common (e.g. Laine et al. 1999 and references therein; Regan \& Mulchaey
1999), and are being found in both Seyferts and in non-active galaxies,
but whether they are more common in the active galaxies, is, for the
time being, an open question. In any case, the radial mass flow they
excite may be too small for fuelling purposes.

\section{Bars and Circumnuclear Star Formation}

Circumnuclear regions (CNRs) of enhanced star formation in barred
galaxies are relatively common, and are believed to occur in response to
gas accumulation between or near ILRs (see
Shlosman 1999). There, gas inflow along the bar is slowed down, and the
resulting high gas density can lead to spectacularly enhanced star
formation. The star formation can result from the spontaneous
gravitational collapse of gas clouds if these become Jeans-unstable
(e.g. Elmegreen 1994), from cloud-cloud collissions, or it can be
triggered by shocks or density-wave spiral armlets (such as in the case
of M100, see Knapen et al. 1995a,b, 1999b).

The importance of star-forming CNRs lies in the fact that they are
excellent inflow laboratories, for a variety of reasons. Firstly, the
phenomena occurring as a result of the inflow are not as violent as in
AGN, where apart from the phenomenology resulting from the inflow
itself, one will also observe secondary effects due to the activity,
such as gas outflow. Secondly, the angular and physical scales in CNRs
are excellent for observational study. At a typical distance of a CNR
host galaxy, 1 arcsec will correspond to up to 100 parsec ($\sim70$ pc
for objects like M100, in the Virgo cluster), which means that
ground-based non-corrected observations can resolve individual starburst
regions and gas flow patterns (e.g. Knapen et al. 1995a, 1999b), whereas
at higher resolution, e.g., using the HST, adaptive optics, or VLBI, one
can resolve the individual star clusters (e.g. Maoz et al. 1995). In
contrast, nuclear starbursts or AGN occur at angular scales much too
small to resolve, e.g., the individual energy sources or powering star
clusters. Thirdly, CNRs are bright across a wide range of 
wavelengths, due to their copious massive star formation activity. Gas
and stars are easy to observe at high resolution in UV (e.g. Maoz et
al. 1995; Colina et al. 1997 using the HST), optical, \ha, NIR
(e.g. Knapen et al. 1995a), mid-IR (e.g. Wozniak et al. 1998),
millimeter (e.g. review by Kenney 1997; Knapen et al. 1999b) and radio
continuum emission (e.g. Hummel, van der Hulst \& Keel 1987).

\subsection{Optical and NIR Observations of CNRs}

Optical and NIR images of ever better quality are being produced,
showing more and more detail in more and more CNRs. The results continue
to comply with the general picture sketched above, but evolutionary
sequences, if any, are still to emerge clearly. The main problem in the
detailed interpretation of optical/NIR imaging is to disentangle the
combined effects of young and old stars, and cold and possibly hot
dust. Multi-color imaging can help, but additional spectroscopy is
necessary to pin down, e.g., populations or ages of the stars
(e.g. Ryder \& Knapen 1999). This area has only recently started to
receive the attention it needs.

\begin{figure}
%\vspace{1cm}
\centering{\epsfxsize=12cm \epsfbox{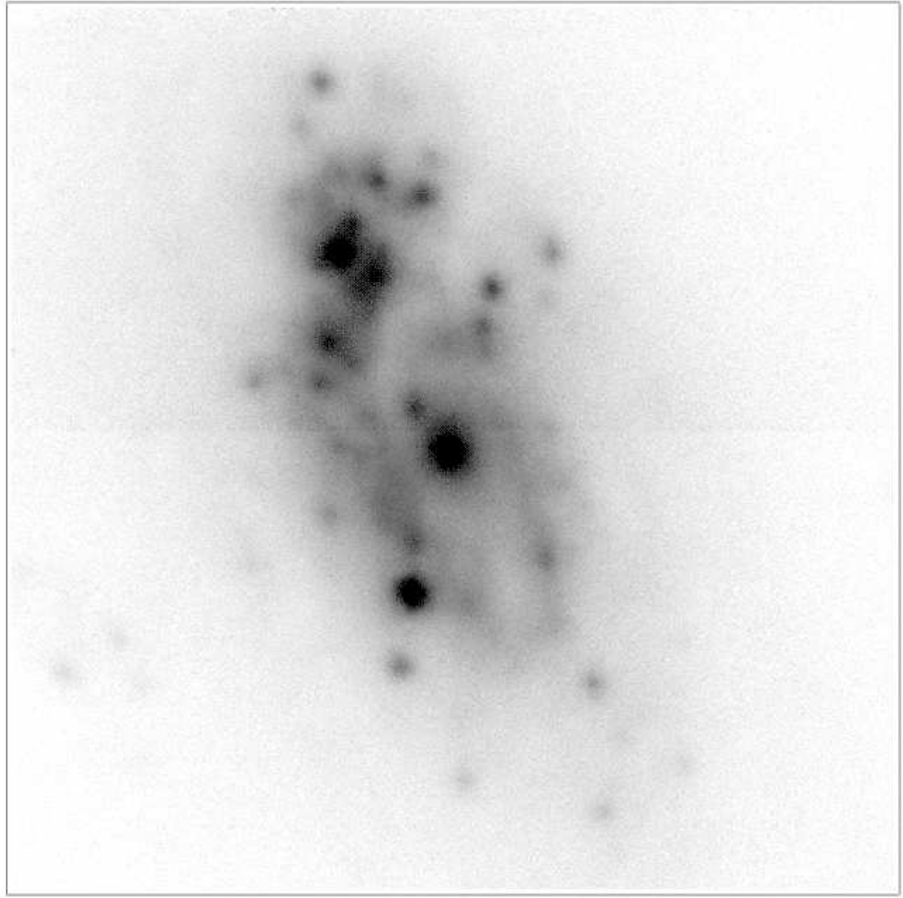}}
\caption{CFHT adaptive optics $K'$-band image of the circumnuclear
star-forming region of the galaxy NGC~2903. The field of view is
21\sec, spatial resolution is estimated to be better than \secd 0.3,
and N is up, E to the left.}
\end{figure}

A major step forward in imaging has occurred thanks to the availability
of high-resolution imaging techniques. In ground-based work, resolutions
of $\sim\secd 0.7$ are standard nowadays (e.g. Elmegreen, Chromey \&
Warren 1998; P\'erez-Ram\'\i rez et al. 1999).  Examples of work based on HST
imaging, and taking advantage of the high resolution it offers, have been
mentioned in the previous section, but especially in the NIR
ground-based telescopes can be competitive. For example, Ryder \& Knapen
(1999) used the upgraded UKIRT to obtain images in the $J, H$ and $K$
bands of the core region of M100 with an angular resolution of
$\sim\secd0.3$, from which over 40 individual emitting knots, presumably
star-forming, could be identified. Adaptive optics (AO) techniques are
also routinely capable of delivering superb resolution, but, until the
advent of laser guide stars, only in selected objects due to the limited
availability of suitable natural sources to correct the wavefront.

As an example, we mention the results obtained by Laine et al. (1999),
who detected a miniature grand-design spiral at scales well below 100 pc
within the CNR of the non-active galaxy NGC~5248. Another example is
shown in Fig.~2, a $K'$ band image at $<\secd 0.3$ resolution of the
circumnuclear region of the (SXT4) galaxy NGC 2903, obtained in Oct.
1998 with the PUEO AO system and the KIR array camera on the
Canada-France-Hawaii Telescope, while the natural, uncorrected, seeing
was about \secd\ 0.6. The image shows a large number of well-resolved
individual knots, generally coinciding with \ha\ emission, and
indicating the important contribution of light from young stars to the
2.2$\mu$m emission. We have observed a small sample of CNRs (Knapen et
al., in preparation), and most of those are similar to NGC~2903 in that
the strong star formation regions in the CNR are obvious in the NIR
emission, confirming that $K$-band emission is not insensitive to a
young stellar population, and that any attempt to derive a gravitational
potential from NIR images must be accompanied by a discussion of a
non-constant $M/L$ ratio (e.g. Knapen et al.  1995a,b, 1999b).

\subsection{Gas Density Observations of CNRs}

Because the \hi\ surface densities in the central regions of barred
galaxies are not high enough to allow high-resolution 21cm observations,
our knowledge of the gas densities in CNRs comes from interferometric
millimeter observations, usually in the CO 1$\rightarrow$0 line. The CO
can be distributed in partial or complete rings, spiral arms, filled
exponential disks, or in ``twin peaks'' (see Kenney 1997 for a
review). In M100, leading spiral arms within the CNR have also been
observed in CO (e.g. Sakamoto et al. 1995; Knapen et al. 1999b). The main
interpretational problem remains the uncertainty in the CO to \h2\
conversion factor, $X$, but it is certain that the standard value cannot
be used without invoking large uncertainties in the derived gas mass
(e.g. Reynaud \& Downes 1999). This problem is less severe in kinematic
studies, although \h2\ present at certain locations in a CNR might not
be accompanied by enough CO to be detected, as a result of a locally very
high value of $X$.

Jogee (1998) studied CO emission from the CNRs in a dozen barred
galaxies, a sample which included starbursts and non-starbursts in order
to study systematic differences between these two classes of
objects. Jogee found that whereas the CO in starbursts reaches larger
peak surface densities, in non-starbursts it is observed to be either
distributed in ring-like structures (e.g. NGC 3351, NGC 4314, NGC 6951),
or extended over the central 1-2 kpc (e.g. NGC 4569, NGC 3359, NGC
7479). The total CO flux (interpreted as gas mass through the use of the
standard $X$ value) is comparable in both classes, though, with peak
molecular gas surface densities 3-4 times higher in the starbursts. This
conclusion does depend on the assumption that the standard value for $X$
can be used.

\subsection{Kinematics of CNRs}

Detailed observation of the gaseous and stellar kinematics is the most
direct way to study the dynamics of barred galaxies, and their central
regions. For large-scale bars, \hi\ interferometry gives the best
measurements of the gas, though at rather low spatial resolution. Two
dimensional stellar kinematics of large-scale bars is harder to obtain
due to the small field of view of the most appropriate instruments,
integral field spectrographs, or the incomplete coverage obtained when
using a collection of long-slit spectra. This area remains much in
need of further observational advance.

\begin{figure}
%\vspace{1cm}
\centering{\epsfxsize=13.5cm \epsfbox{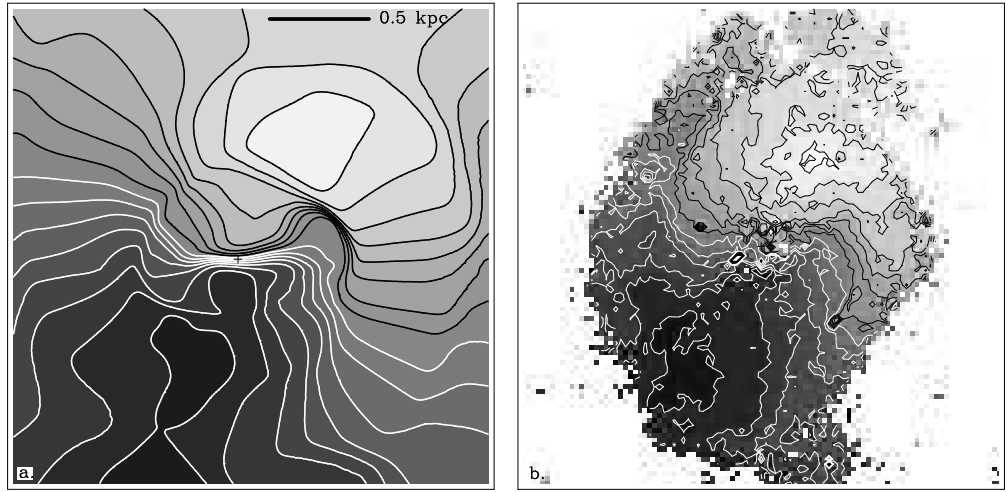}}
\caption{Left panel ({\it a.}): Gas velocity field for the CNR in M100
as derived from the SPH model of Knapen et al. (1995b). Contour
separation is 15\kms, and the scale is indicated in the top right hand
corner. Right panel ({\it b.}): TAURUS Fabry-P\'erot \ha\ velocity
field at $\sim\secd 0.7$ resolution 
obtained with the 4.2m William Herschel Telescope, shown at 
the same scale and orientation
as the model velocity field. The region shown is
29\sec\ across. N is up, E to the left.}
\end{figure}

For the CNRs the prospects for observing high-resolution velocity fields
are much better, due to the higher flux densities in the main tracers,
and the smaller angular size of the area of interest. For the gas, CO
interferometry now gives data on gas densities and kinematics in CNRs at
spatial resolutions as good as \secd 1.6 (Reynaud \& Downes 1999), while
improvements to telescopes will continue to improve this resolution
limit. Ionized gas can be observed with e.g.  Fabry-P\'erot
spectroscopic imaging in the \ha\ line, which reaches sub-arcsec
resolution over a substantial field of view (e.g. Knapen et al. 1999b;
Fig.~3).

Integral field spectrographs, which project individual small regions
(comparable to pixels or small beams) of a galaxy image onto a
spectrograph, are powerful instruments due to the high resolution
achievable and the possibility to make multi-line measurements (giving,
e.g., stellar and gaseous kinematics), but the problem remains the
limited number of individual regions that can be observed in an
exposure, which translates most directly into a compromise between
spatial resolution and field of view. To reach well-sampled sub-arcsec
resolution the field usually remains limited to around 10\sec\
(e.g. Arribas et al. 1999; Colina \& Arribas 1999; Garc\'\i a-Lorenzo,
Mediavilla \& Arribas 1999; Miller et al. 1999).

Kinematic results for star-forming CNRs usually show a rather
unperturbed ``spider diagram'' in the velocity field, indicating
predominant disk rotation (e.g. NGC 3351 and NGC 4314: Jogee 1998; NGC
985: Arribas et al. 1999; NGC 5248: Laine et al. 1999b).  In some cases
there are marked deviations from circular motion, due to gas streaming
across miniature spiral armlets in the CNR, and/or along the inner bar
(e.g. M100: Knapen et al. 1999b; Fig.~3). However, such non-circular
motions are only observable if the position angles of the galaxy's
kinematic axis and of the kinematic structures in the CNR are favorably
different, and they may only be observable with high-resolution
observations due to the small angular scales of most CNRs. Thus, more
and better observations are needed to come to more general conclusions
as to how common (or uncommon) predominant circular rotation is in CNRs,
and what the consequences are for our understanding of the evolution of
CNRs and for star formation processes within them.

Qualitative and quantitative comparison of kinematic data with numerical
modeling is a very interesting area which should allow clearer
discrimination between models than the usually performed morphological
comparison. Kinematic comparison, however, is not often done, even in
cases where kinematic data is available, such as in most studies where
modeling is based on interferometric CO data. For M100, we found good
agreement (Knapen et al. 1999b; Fig.~3) between the main kinematic
features, such as rotation curve shape, and deviations in the velocity
field due to spiral arm or bar streaming motions, as observed in \ha\
and CO, and as obtained from an analysis of our SPH model of the CNR in
this galaxy (Knapen et al. 1995b). This qualitative and quantitative
agreement forms an important confirmation of the modeling and
interpretation in terms of a resonant structure driven by the moderately
strong stellar bar.

\section{Bars at Cosmological Distances}

Studying bar fractions, and perhaps even properties, at cosmological
distances can in principle yield direct evidence for or against galaxy
formation and evolution models. One such model directly relevant to the
topic in hand is that of secular evolution, where bulges are thought to
form through the collapse of bars.  Better datasets, in particular the
Hubble Deep Fields (HDFs; Williams et al. 1996), make high-quality
imaging of large samples of galaxies at $0<z<2$ available.  Among the
problems that need to be addressed is that of bandshifting: at
$z\approx2$, we see $U$ light redshifted into $I$; at $z\approx4$,
far-UV shifts into the $I$-band. Can redshifted barred galaxies be
reliably recognized as such?

The results are very limited so far, but they are exciting and warrant
substantial further research. The bar fraction has not been systematically
studied in the Medium Deep Survey, but at least some barred galaxies are
present (Abraham et al. 1996).  Although several barred spirals can be
recognized in the northern HDF, van den Bergh et al. (1996) only find
one bona fide case at $I>21$mag (corresponding to a bar fraction of
0.3\%), and 6 possible ones, as based on a visual inspection of all
galaxies. As pointed out by Abraham et al. (1999), however, this result
remains controversial because barred galaxies may be undetectable at the
levels studied by van den Bergh et al. (1996), due to the combined
effects of bandshifting and low signal-to-noise.

From a study of the two HDFs (N and S), Abraham et al. (1999) are able
to compile a sample of 18 barred galaxies with low inclination, selected
on the basis of their axis ratios and isophote twists. The
$z$-distribution of this sample supports the conclusion reached by van
den Bergh et al. (1996) of a significant decrease of the fraction of
barred spirals beyond redshift of $z=0.5$. Abraham et al. (1999) briefly
discuss a number of possibilities responsible for this decrease, but do
not reach firm conclusions. Further study is required into observational
issues like the amounts of dust extinction as a function of redshift,
which possibly changes the morphological appearance of galaxies, or
into the details of the bar detection mechanisms. Local studies have
shown the difficulties in objectively defining whether a galaxy is
barred (see Sect.~4), and classification of galaxies at substantial
redshifts might well be biased to those galaxies that we would call
strongly, or obviously, barred at $z\sim0$.

\section{Summary and Open Questions}

Barred galaxies are plentiful, and important for a variety of issues,
ranging from star formation to central activity and galaxy evolution.
There is an enormous amount of observational and theoretical work on
barred galaxies, and this review has attempted to present some of the
exciting progress being made.

However, there are still many outstanding issues. In this section we
will mention a few of the areas where observational progress can be
made, leading to a better understanding of bars in galaxies, their
origin and formation, and their relation to their host galaxies and the
central or circumnuclear activity that may take place in their core
regions.

\begin{itemize}

\item Most systematic studies of properties of barred galaxies, based on
large samples, are still limited because of the use of catalogues based
on blue or visual plates, and with poor resolution. There is a need not
only for a NIR all sky survey at decent and uniform resolution and
limiting brightness levels, but also of a reliable catalogue based upon
it. Projects such as the ongoing 2MASS surveys or planned survey
telescopes (e.g. the Visible and Infrared Survey Telescope for Astronomy
- VISTA) should lead to progress in this area.
 
\item Gaseous kinematics of bars has been observed for a few decades
now, and is reasonably well understood. Using \ha\ or CO observations,
spatial resolution has reached comfortably low levels of near or below 1
arcsec, while upgrades to existing array telescopes and especially the
planned Atacama Large Milimetre Array (ALMA) will allow much better
resolutions. The area of stellar kinematics, however, needs substantial
attention. Long-slit spectra are available but cannot show details of
two-dimensional velocity fields. Integral field spectrographs are
possibly the way forward, but the field of view is usually still too
limited to study large-scale bar kinematics in sufficient detail.

\item It has now been established theoretically, as well as
observationally, that gas can flow inwards in barred galaxies, toward
the active, starbursting, or star forming central regions. But we need
to study the detailed interrelations between the dynamics of the bar and
central regions on the one hand, and the star formation resulting from
it, but also influencing the kinematics, on the other. Kinematic
observations of gas and stars, and NIR imaging and spectroscopy, all at
high-resolution, in combination with detailed numerical modelling, are
required.

\item The study of bar fractions at moderate and high redshift is an
emerging area that needs further attention. It can in principle provide
important observational inputs and checks to various models of galaxy,
bar and bulge formation and evolution, but the various problems related
to, e.g., identification and completeness require continued scrutiny.

\end{itemize}

{\bf Acknowledgments} I am indebted to my many collaborators on various
programs, especially John Beckman and Isaac Shlosman.  I thank Shardha
Jogee, Seppo Laine, Mike Merrifield, Isaac Shlosman, Nial Tanvir, and
Jeremy Yates for comments on an earlier version of the manuscript. The
Canada-France-Hawaii Telescope is operated by the National Research
Council of Canada, the Centre National de la Recherche Scientifique de
France and the University of Hawaii. Partly based on observations
obtained at the William Herschel Telescope, operated on the island of La
Palma by the Royal Greenwich Observatory in the Spanish Observatorio del
Roque de los Muchachos of the Instituto de Astrof\'\i sica de Canarias.

\end{document}